\documentclass[12pt]{article}
\usepackage{a4wide}
\usepackage{amssymb}
\usepackage{xcolor}
\begin{document}
{\renewcommand{\thefootnote}{\fnsymbol{footnote}}
\begin{center}
{\LARGE  Hypersurface-deformation algebroids\\ and effective space-time models}\\
\vspace{1.5em}
Martin Bojowald,$^1$\footnote{e-mail address: {\tt bojowald@gravity.psu.edu}}
Suddhasattwa Brahma,$^{1,2}$\footnote{e-mail address: {\tt suddhasattwa.brahma@gmail.com}} Umut
B\"{u}y\"{u}k\c{c}am,$^1$\footnote{e-mail address: {\tt uxb101@psu.edu}} \\
and Fabio D'Ambrosio$^{1,3,4}$
\\
\vspace{0.5em}
$^1$ Institute for Gravitation and the Cosmos,\\
The Pennsylvania State
University,\\
104 Davey Lab, University Park, PA 16802, USA\\
\vspace{0.5em}
$^2$ Center for Field Theory and Particle Physics,\\
Fudan University, 200433 Shanghai, China\\
\vspace{0.5em}
$^3$ Institut f\"ur Theoretische Physik, ETH-H\"onggerberg, CH-8093 Z\"urich\\
\vspace{0.5em}
$^4$ CPT, Aix-Marseille Universit\'e, Universit\'e de Toulon, CNRS,
Case 907,\\ F-13288 Marseille, France\\
\vspace{1.5em}
\end{center}
}

\setcounter{footnote}{0}

\newtheorem{theo}{Theorem}
\newtheorem{prop}{Proposition}
\newtheorem{lemma}{Lemma}
\newtheorem{defi}{Definition}

\newcommand{\proofend}{\raisebox{1.3mm}{\fbox{\begin{minipage}[b][0cm][b]{0cm}
\end{minipage}}}}
\newenvironment{proof}{\noindent{\it Proof:} }{\mbox{}\hfill \proofend\\\mbox{}}
\newenvironment{ex}{\noindent{\it Example:} }{\medskip}
\newenvironment{rem}{\noindent{\it Remark:} }{\medskip}

\newcommand{\case}[2]{{\textstyle \frac{#1}{#2}}}
\newcommand{\lP}{\ell_{\mathrm P}}

\newcommand{\md}{{\mathrm{d}}}
\newcommand{\tr}{\mathop{\mathrm{tr}}}
\newcommand{\sgn}{\mathop{\mathrm{sgn}}}

\newcommand*{\R}{{\mathbb R}}
\newcommand*{\N}{{\mathbb N}}
\newcommand*{\Z}{{\mathbb Z}}
\newcommand*{\Q}{{\mathbb Q}}
\newcommand*{\C}{{\mathbb C}}

\begin{abstract}
  In canonical gravity, covariance is implemented by brackets of
  hypersurface-deformation generators forming a Lie algebroid. Lie algebroid
  morphisms therefore allow one to relate different versions of the brackets
  that correspond to the same space-time structure.  An application to
  examples of modified brackets found mainly in models of loop quantum gravity
  can in some cases map the space-time structure back to the classical
  Riemannian form after a field redefinition. For one type of quantum
  corrections (holonomies), signature change appears to be a generic feature
  of effective space-time, and is shown here to be a new quantum space-time
  phenomenon which cannot be mapped to an equivalent classical structure. In
  low-curvature regimes, our constructions prove the existence of classical
  space-time structures assumed elsewhere in models of loop quantum cosmology,
  but also shows the existence of additional quantum corrections that have not
  always been included.
\end{abstract}

\section{Introduction}

Several independent examples of modified gauge transformations have been found
in different models of canonical quantum gravity, using effective
\cite{ConstraintAlgebra,JR,ModCollapse,LTBII,ScalarHol,ScalarTensorHol,HigherSpatial}
and operator calculations
\cite{ThreeDeform,TwoPlusOneDef,TwoPlusOneDef2,AnoFreeWeak,SphSymmOp}. In
classical canonical formulations, space-time structure is encoded not in the
usual form of general covariance of tensors, but by the equivalent version of
gauge covariance under hypersurface deformations in space-time
\cite{DiracHamGR,Regained}. The new structures found as a direct consequence
of key ingredients of the quantization process using holonomies instead of
connections therefore confirm a general expectation: Quantum geometry may lead
to modified space-time structures \cite{Action,DeformedRel}.

Although these modified gauge structures have been found within a variety of
models of loop quantum gravity and by virtue of different computational
methods, they all share some important properties. There is a phase-space
function $\beta$ modifying only the Poisson bracket of two smeared
Hamiltonian constraints (or normal deformations of hypersurfaces). Denoting
the constraints by $H[N]$ with the lapse function $N$ that specifies the
magnitude of the normal deformation at every point on a spatial hypersurface,
we have
\begin{equation}\label{HHbeta}
\left\{H[N],H[M]\right\} = -H_{a}[\beta q^{ab}(N \partial_bM-M \partial_bN)]\,.
\end{equation}
On the right-hand side, $H_a$ are the components of the diffeomorphism
constraint  (generating tangential deformations) and $q^{ab}$ is the inverse
metric on a spatial hypersurface. Brackets involving $H_a[M^a]$ retain the
classical form
\begin{eqnarray}
 \{H_a[M_1^a],H_b[M_2^b]\} &=& -H_c[{\cal L}_{M_2}M_1^c] \label{DD}\\
 \{H[N],H_a[M^a]\} &=& -H[{\cal L}_MN]\,. \label{HD}
\end{eqnarray}
There have been attempts to modify the brackets involving not only the
Hamiltonian constraint as in (\ref{HHbeta}) but also the diffeomorphism
constraint \cite{VectorHol,DiffeoOp}. Other such examples are given by
fractional space-time models, in which the modification functions can,
however, be absorbed \cite{FractHypDef}.  A discrete version of the brackets
has been defined in \cite{DiscDirac}, which differs from (\ref{DD}) and
(\ref{HD}).  In the present paper, we focus on continuum effective theories in
which space (but not necessarily space-time) has the classical
structure. Accordingly, (\ref{DD}) will not be modified.  We will derive a new
form of brackets in which (\ref{HD}) is modified, but (\ref{DD}) is
not. Nevertheless, our main focus will be on brackets with modifications as in
(\ref{HHbeta}).

The correction function $\beta\neq 1$ depends on the phase-space variables,
and transforms as a spatial scalar. In the classical case, the
hypersurface-deformation brackets are (on shell) related to the Lie algebra of
space-time diffeomorphims, reflecting the coordinate invariance of general
relativity.  Brackets with $\beta\not=1$ modify general covariance of the
effective theory, but in such a way that no gauge transformations are
violated. (Obeying the condition of anomaly freedom, gauge transformations are
allowed to be modified by quantum corrections but not to be destroyed.)

With modified brackets, the effective metric $q_{ab}$ appearing in
(\ref{HHbeta}) cannot be part of a space-time line element of classical form:
Modified gauge transformations of $q_{ab}$, generated by $H[\epsilon]$ and
$H_a[\epsilon^a]$, do not complement coordinate transformations of ${\rm
  d}x^a$ to form an invariant space-time line element
\begin{equation}\label{ds}
 {\rm d}s^2=-N^2{\rm
  d}t^2+q_{ab}({\rm d}x^a+N^a{\rm d}t)({\rm d}x^b+N^b{\rm d}t)
\end{equation}
in canonical form. Nevertheless, there may be field redefinitions of different
kinds which allow one to find a classical space-time picture for some function
$\beta$ of the phase-space variables. For instance, in some cases $\beta$ can
be absorbed in the lapse function by $N':=\sqrt{|\beta|}N$, with classical
brackets in terms of $N'$.  Or, a combination of the original spatial metric
and extrinsic curvature could determine the spatial geometry of an effective
space-time of classical type.  The question has been investigated in certain
spherically symmetric models in \cite{Absorb}, with some encouraging results:
Gauge transformations of the original canonical fields of the effective theory
(including $q_{ab}$) are deformed, but by applying canonical transformations
it is possible, in some cases, to recover the classical
hypersurface-deformation brackets and hence to restore general
covariance. Specifically, a canonical transformation with this effect has been
found in \cite{Absorb} when $\beta$ depends only on metric components.
Absorbing $\beta$ in $q^{ab}$ then provides a simple canonical
transformation. If $\beta$ depends on the momentum (extrinsic curvature) as
well, it is more difficult to see whether it can be removed from the brackets.

In the present paper, we analyze the same question from a different
perspective which is insensitive to the availability of canonical
transformations. Our discussion makes use of the general setting of Lie
algebroids, of which a suitable fiber-bundle formulation of (\ref{HHbeta})
provides an example \cite{ConsAlgebroid}. More generally, the language of Lie
algebroids is a well-defined mathematical structure that allows one to
formalize theories with structure functions.  Our results are independent of
details of any specific form of quantum gravity in the sense that we will not
use equations or methods characteristic of a specific approach. Instead, we
use the general form (\ref{HHbeta}) of the modified bracket of two normal
deformations as a guiding principle and study possible Lie-algebroid
realizations.  Modifications of the classical brackets can be understood as a
generic form of quantum corrections, introduced by some effective quantum
gravity theory.

We will be able to classify different inequivalent space-time structures
corresponding to modified brackets of the type (\ref{HHbeta}) that cannot be
related by morphisms. While there appears to be an arbitrary modification
function $\beta$ in (\ref{HHbeta}) with virtually unrestricted quantum
corrections, only ${\rm sgn}\beta$ remains as the single choice left after
equivalence classes of brackets up to morphisms are considered.  This result
helps to clarify the implications of modified brackets (\ref{HHbeta}) for
space-time structures. In particular, they can be related to the classical
brackets by Lie algebroid morphisms as long as $\beta$ has a definite sign and
is non-zero. The existence of effective Riemannian space-time structures is
confirmed in this case, which so far has only been assumed, for instance in
\cite{Hybrid,QFTCosmo,QFTCosmoClass,AAN}. Such modifications therefore do not
imply radical changes of the space-time structure, even though they may still
lead to a modified dynamics on and of the effective space-time. If $\beta$
does not have a fixed sign, a new version of quantum space-time is obtained
which exhibits signature change as a new physical effect.

In some cases, concrete morphisms can be formulated with simple
interpretations of their implications on canonical variables and the
dynamics. For instance, with spatially constant $\beta\not=1$, as in
cosmological models with first-order perturbative inhomogeneity, a suitable
morphism is obtained by changing the usual conventions in setting up the
canonical formulation based on space-time foliations into spatial
slices. Somewhat akin to absorbing $\beta$ in the lapse function, one can make
use of a generalized canonical formulation which is a hybrid version of, on
one side, Dirac's \cite{DiracHamGR} and the ADM \cite{ADM} formulation with
variables adapted to directions normal and tangent to a spatial hypersurface,
and on the other Rosenfeld's \cite{Rosenfeld} earlier derivation of canonical
gravity without reference to a foliation or preferred directions.  We will use
a foliation, but do not require the timelike vector $n^{\mu}$ to be normalized
or orthogonal to the spatial tangent plane.  The normalization function
$n^{\mu}n_{\mu}$ can be related to $\beta$. Therefore, non-standard
normalizations present a more-general way of relating modified brackets to
classical space-time structures than absorbing $\beta$ in the lapse function
would do.  The angles between $n^{\mu}$ and the spatial tangent plane give
rise to new modifications of the brackets not yet encountered elsewhere. At
the same time, we make use of a concise derivation of the
hypersurface-deformation brackets and use the example to introduce Lie
algebroids in this context. Morphisms of Lie algebroids will lead to further
transformations that can be used to relate modified brackets of different
types, still with the classical signature as the only parameter that
characterizes inequivalent space-time structures of brackets of the form
(\ref{HHbeta}) via ${\rm sgn}\beta$. This result allows us to draw rather
general conclusions about implications of the modified dynamics according to
(\ref{HHbeta}).

\section{Canonical gravity and Lie algebroids}
\label{sec:1.2}

In order to set up the canonical formalism, we assume, as usual, space-time
${\cal M}$ to be globally hyperbolic and introduce a foliation by
constant-level surfaces of a parameter $t\in\mathbb{R}$, such that the
hypersurfaces are all spacelike.  Each spatial slice is homeomorphic to a
3-manifold $\sigma$, on which we may choose local coordinates $x^{a}$, $a\in
\{1,2,3\}$. We realize $\sigma$ as a spatial hypersurface
$\Sigma_t:=X_t(\sigma)$ at constant $t$ by an embedding $X\colon
\mathbb{R}\times\sigma \hookrightarrow \mathcal{M}$ with $(t,x) \mapsto
X(t,x)=: X_t(x)$.

We choose a foliation $X_t=X(t,\cdot)$ and define a time-evolution vector
field $\tau^{\mu}$ by
\begin{equation}\label{tau}
\tau(X) := \partial_t X^{\mu}(t,x)\partial_{\mu}\,.
\label{eq:Tau}
\end{equation}
This vector field is, in general, not normal to $\Sigma_t$. Following ADM
\cite{ADM}, it is convenient to introduce vector fields tangential to
$\Sigma_t$, given by
\begin{equation}\label{S_a}
X_a(X) := \partial_a X^{\mu}(t,x)\partial_{\mu}\,,
\label{eq:Sa}
\end{equation}
and to define a time-like vector field normal to the time slice $\Sigma_t$
by
\begin{equation}\label{n}
g_{\mu\nu}n^{\mu}X_a^{\nu} = 0\quad,\quad g_{\mu\nu}n^{\mu}n^{\nu} =-1\,.
\end{equation}
If we further require that $n^{\mu}$ point toward the future, that is,
$n^{\mu}\partial_{\mu}t>0$, it is uniquely defined.  By introducing the lapse
function $N(X)$ and the shift vector field $M^{a}(X)$ the time-evolution
vector field $\tau^{\mu}$ is decomposed into its components normal and
tangential to $\Sigma_t$:
\begin{equation}
\tau^{\mu}(X) = N(X)n^{\mu}(X) + N^{a}(X)X_a^{\mu}(X)\,.
\label{eq:taudecomp}
\end{equation}
Since the choice of the embedding $X$ is arbitrary, the components of lapse
and shift are free functions as long as they give rise to a timelike
$\tau^{\mu}$.

So far, we have used only well-known and basic ingredients of the canonical
formulation. (See \cite{CUP} for further details.) The decomposition
(\ref{eq:taudecomp}) and the normalization condition of $n^{\mu}$ in (\ref{n})
play a key role in our considerations of modified space-time structures. In
order to exhibit the full freedom of the formalism, we will not follow the
common convention of normalizing $n^{\mu}$ by
$g_{\mu\nu}n^{\mu}n^{\nu}=-1$. We may fix any other negative constant, or even
a phase-space function, for Lorentzian space-time signature, or a positive
constant (or phase-space function) for Euclidean signature. We may therefore
require that $g_{\mu\nu}n^{\mu}n^{\nu}=\epsilon\beta$, where $\epsilon=-1$ in
the Lorentzian case and $\epsilon=+1$ in the Euclidean case. If the signature
is constant, $\beta>0$ is a positive phase-space function. But in anticipation
of applying these methods to some of the models found in the context of loop
quantum gravity, we allow for $\beta$ to change its sign, so that ${\rm
  sgn}\beta=:\epsilon_{\beta}$ may not be constant. The overall signature is
then locally given by the product $\epsilon\epsilon_{\beta}$.

In order to compare dynamical results obtained with different normalizations,
we should demand that $\tau^{\mu}(X)$ remain the same and be independent of
$\beta$:
\begin{equation}
  \tau^{\mu}(X) = \frac{1}{\sqrt{|\beta|}}N(X)n^{\mu}(X) + M^{a}(X)X_a^{\mu}(X)=:
  N_{\beta}(X)n^{\mu}(X) + M^{a}(X)X_a^{\mu}(X)
\label{eq:taudecompmod}
\end{equation}
where now $n^{\mu}/\sqrt{|\beta|}$ is normalized to
$\pm1=\epsilon\epsilon_{\beta}$.  This condition ensures that equations of
motion for evolution along $\tau^{\mu}$ exist independently of the canonical
decomposition in terms of hypersurfaces. At this stage, we see the simple
result that the lapse function has to absorb any non-standard normalization
factor $\beta$, but later on we will be able to draw more benefit from these
simple-looking considerations. The only requirement for
(\ref{eq:taudecompmod}) to be used is that $n^{\mu}$ and
$M^{\mu}=M^aX_a^{\mu}$ form a basis of the tangent space to ${\cal M}$ at each
point. We may therefore drop normalization conditions as well as orthogonality
of $n^{\mu}$ and $M^{\mu}$.

\subsection{A concise derivation of the hypersurface-deformation brackets}
\label{s:brief}

We derive the brackets of hypersurface deformations with non-standard
normalization by repurposing a derivation of the usual result given in
\cite{ConsAlgebroid}. The main aim of this paper was to analyze the
Lie-algebroid structure of the brackets, which we will describe in the
following subsection. Some part of the mathematical analysis of
\cite{ConsAlgebroid} amounts to a brief derivation of the brackets which we
formulate here in abstract index notation and, at the same time, use it to
derive the brackets with non-standard normalization. As a further
generalization, we will also assume a non-orthogonality relation between
$n^{\mu}$ and $X_a^{\mu}$. More traditional derivations using ADM-style
evolution equations or geometrodynamics are given in App.~\ref{a:ADM} for the
case of a non-unit normal $n^{\mu}$, with equivalent results.

The explicit derivation of hypersurface deformations depends on
choices of coordinates or embedding functions, but the brackets must be
covariant under changes of these auxiliary structures. As in
\cite{ConsAlgebroid}, one can exploit the coordinate freedom by working
with embeddings such that the space-time metric, from which the spatial metric
$q_{ab}$ in the structure functions is induced, is Gaussian with respect to
the hypersurfaces:
\begin{equation} \label{Gauss}
{\rm  d}s^2=\epsilon {\rm d}t^2+q_{ab}{\rm d}x^a{\rm d}x^b\,.
\end{equation}
In this way, one fixes a representative in each equivalence class of
hypersurface embeddings. The remaining coordinate freedom is given by
diffeomorphisms generated by so-called $g$-Gaussian vector fields $v^{\mu}$
which preserve the Gaussian form of the metric and therefore satisfy
\begin{equation} \label{gGauss}
 n^{\mu}{\cal L}_vg_{\mu\nu}=0
\end{equation}
with some vector field $n^{\mu}$ normal to $t={\rm constant}$, but not
necessarily normalized.  This condition ensures that an infinitesimal
diffeomorphism along $v^{\mu}$, changing $g_{\mu\nu}$ to
$g'_{\mu\nu}:=g_{\mu\nu}+{\cal L}_vg_{\mu\nu}$, respects the relations
$n^{\mu}n^{\nu}g'_{\mu\nu}=n^{\mu}n^{\nu}g_{\mu\nu}= \epsilon$ and
$n^{\mu}w^{\nu}g'_{\mu\nu}=0$ if $n^{\mu}w^{\nu}g_{\mu\nu}=0$ of the Gaussian
system.  Because they generate diffeomorphisms preserving the Gaussian form
of the metric, $g$-Gaussian vector fields form a subalgebra of the Lie algebra
of all vector fields with bracket the usual Lie bracket. As found in
\cite{ConsAlgebroid}, one can derive the hypersurface-deformation brackets by
rewriting the Lie bracket using properties of vector fields $v^{\mu}$
satisfying (\ref{gGauss}).

Some restriction on the form of vector fields is necessary because the
hypersurface deformations as gauge transformations are known to be equal to
infinitesimal space-time diffeomorphisms only on-shell \cite{Regained}, that
is, when some of the generators $H$ and $H_a$ and the equations of motion they
generate are set to zero as phase-space functions. The restriction is
implemented here by using $g$-Gaussian vector fields, which turn out to have
Lie brackets directly related to the hypersurface-deformation brackets. Such a
restriction cannot be chosen arbitrarily but must fulfill three conditions:
(i) The vector fields considered must provide a unique extension from spatial
(lapse) functions $N$ and spatial (shift) vector fields $M^a$ to a space-time
vector field $v^{\mu}$ which equals $N n^{\mu}+M^aX_a^{\mu}$ on the spatial
slice. If this condition is fulfilled, it is possible to compute space-time
Lie brackets. (ii) The vector fields considered must form a subalgebra of the
Lie algebra of all space-time vector fields. And (iii), the Lie bracket of
space-time extensions of two pairs $(N_1,M_1^a)$ and $(N_2,M_2^a)$ should not
depend on the extensions but only on spatial derivatives of $N_i$ and $M_i^a$
in addition to the functions and vector fields themselves. With conditions
(ii) and (iii) fulfilled, it is then possible to interpret the Lie bracket of
extensions of two pairs $(N_1,M_1^a)$ and $(N_2,M_2^a)$ as the unique
extension of a third pair $(N_3,M_3^a)$, and to define a new bracket
$[(N_1,M_1^a), (N_2,M_2^a)]:=(N_3,M_3^a)$. All three conditions can be shown
to be true for $g$-Gaussian vector fields \cite{ConsAlgebroid}, recovered as a
special case ($\beta=1$ and $\alpha^a=0$) of the following calculations. To
the best of our knowledge, it is not known whether $g$-Gaussian vector fields
are the only choice fulfilling all three conditions, but having one such
choice is sufficient for a derivation of the brackets.

We first derive properties (i), (ii) and (iii), found in \cite{ConsAlgebroid},
using abstract index notation. We write (\ref{gGauss}) as
\begin{equation}
 0= n^{\mu}{\cal L}_vg_{\mu\nu}= n^{\mu}v^{\rho}\partial_{\rho}g_{\mu\nu}+
 n^{\mu}g_{\nu\rho}\partial_{\mu}v^{\rho}+
 n^{\mu}g_{\mu\rho} \partial_{\nu}v^{\rho}\,.
\end{equation}
The first two terms can be expressed by the Lie bracket of $n^{\mu}$ and
$v^{\nu}$ if we write $n^{\mu}v^{\rho}\partial_{\rho}g_{\mu\nu}=
v^{\rho}\partial_{\rho}n_{\nu}- g_{\mu\nu}v^{\rho}\partial_{\rho}n^{\mu}$. The
last term in $n^{\mu}{\cal L}_vg_{\mu\nu}$ can be replaced by a total
derivative using
$n^{\mu}g_{\mu\rho}\partial_{\nu}v^{\rho}= \partial_{\nu}(n^{\mu}v^{\rho}g_{\mu\rho})-
v^{\rho}\partial_{\nu}n_{\rho}$. In addition to the Lie bracket and the total
derivative, there remain two extra terms related to the 2-form ${\rm d}n$:
\begin{equation}
 0= n^{\mu}{\cal L}_vg_{\mu\nu}
= [n,v]^{\mu}g_{\mu\nu}+\partial_{\nu}(n^{\mu}v^{\rho}g_{\mu\rho})+
v^{\rho}({\rm d}n)_{\rho\nu}\,. \label{gGaussd}
\end{equation}

If $n^{\mu}$ is hypersurface orthogonal, by the Frobenius theorem we have
${\rm d}n=n\wedge w$ with some 1-form $w$ which can, without loss of
generality, be assumed to be orthogonal to $n^{\mu}$. For
$n^{\mu}n_{\mu}=\epsilon$ and the metric in Gaussian form, $w=0$ because
$n=\epsilon{\rm d}t$ is closed. In this case, $n^{\mu}$ is hypersurface
orthogonal in a neighborhood of the initial slice by construction of the
Gaussian system. If $n^{\mu}n_{\mu}=\epsilon\beta$, the analog of the Gaussian
system has $n^{\mu}$ hypersurface orthogonal only if $\beta$ is spatially
constant. In order to allow for spatially non-constant $\beta$, we use a
Gaussian system constructed from a unit normal, which would be
$\tilde{n}^{\mu}:=n^{\mu}/\sqrt{|\beta|}$ if
$n^{\mu}n_{\mu}=\epsilon\beta$. This rescaled normal is extended to a closed
1-form in its Gaussian system. We can compute ${\rm d}n=n\wedge w$ from the
equation ${\rm d}\tilde{n}=0$, resulting in $w=-\frac{1}{2}\beta^{-1}({\rm
  d}\beta -\epsilon|\beta|^{-1/2}\dot{\beta} n)$. The second term, with
$\dot{\beta}=\partial\beta/\partial t$, is chosen such that
$n^{\mu}w_{\mu}=0$.

We include one further generalization by relaxing the usual orthogonality
relation to $g_{\mu\nu}n^{\mu}X_{a}^{\nu} = \alpha_{a}$ with fixed phase-space
functions $\alpha_a$ allowed to be non-zero. The components of $\alpha_a$ are
related to the direction cosines (hyperbolicus) of $n^{\mu}$ with respect to
the spatial basis $X_a^{\nu}$. The new condition can equivalently be written
as an orthogonality relation $g_{\mu\nu}n'^{\mu}X_{a}^{\nu} = 0$ with a
redefined $n'^{\mu} := n^{\mu} - \alpha^{a}X^{\mu}_{a}$. With the non-standard
normalization of $n^{\mu}$, the redefined vector satisfies
$n'^{\mu}n'_{\mu}=\epsilon\beta-\alpha_a \alpha^a =: \epsilon\gamma$. In the
Euclidean case, $\epsilon=1$, we must have $\gamma>0$ and therefore
$\alpha^a\alpha_a<\beta$. The same condition ensures that $n^{\mu}$ and
$X_a^{\mu}$ form a basis because the angle between the direction $n^{\mu}$ and
the spatial tangent plane spanned by $X_a^{\mu}$ is less than ninety
degrees. In the Lorentzian case, $\alpha^a\alpha_a$ is unrestricted.

We construct a Gaussian system as before.  The hypersurface orthogonal vector
is now given by
$\tilde{n}'^{\mu}:=n'^{\mu}/\sqrt{|\beta-\epsilon\alpha^a\alpha_a|}=
n'{}^{\mu}/\sqrt{|\gamma|}$. Computing ${\rm d}n'=n'\wedge w$ from the
equation ${\rm d}\tilde{n}'=0$ now results in $w=-\frac{1}{2}\gamma^{-1}({\rm
  d}\gamma- \epsilon|\gamma|^{-1/2}\dot{\gamma}n')$.  With the redefined
normal, (\ref{gGaussd}) takes the form
\begin{equation} \label{gGaussprime}
 0=n'^{\mu}{\cal
   L}_vg_{\mu\nu}=[n',v]^{\mu}g_{\mu\nu}+\partial_{\nu}(n'^{\mu}v^{\rho}g_{\mu\rho})+
v^{\rho}({\rm d}n')_{\rho\nu}\,.
\end{equation}
We use $n'^{\mu}$ because we need a normal vector for the condition of a
$g$-Gaussian vector field. However, we may decompose a $g$-Gaussian vector
field $v^{\mu}$
according to our original basis $(n^{\mu},X_a^{\nu})$ or according to the
redefined basis using $n'^{\mu}$ instead of $n^{\mu}$:
\begin{equation}
 v^{\mu}=N n^{\mu}+M^{\mu}= N n'^{\mu}+ M'^{\mu}
\end{equation}
with $M^{\mu}=M^aX_a^{\mu}$ and $M'^{\mu}=M^{\mu}+N\alpha^aX_a^{\mu}$, or
$M'^{a}=M^a+N\alpha^a$. The latter choice simplifies some derivations and is
therefore employed below, but for full generality we will transform the final
result to a decomposition with respect to $(n^{\mu},X_a^{\nu})$.

We will need the following ingredients in order to rewrite (\ref{gGaussprime})
with a decomposed vector field $v^{\mu}$.  In contrast to the standard case,
$n'^{\mu}n'_{\mu}=\epsilon\gamma$ is not a constant because $\alpha^a$ and
$\beta$ may depend on space and time via phase-space variables. Therefore, for
spatial $M^{\mu}$ (or $M'^{\mu}$), $[n',M]^{\mu}$ has a normal component given
by
\begin{eqnarray}
 \frac{n'^{\mu}n'_{\nu}[n',M]^{\nu}}{n'^{\kappa}n'_{\kappa}} &=&
 \frac{1}{\epsilon\gamma}
 n'^{\mu} \left(n'_{\nu}n'^{\rho} \nabla_{\rho}M^{\nu}-
   n'_{\nu}M^{\rho}\nabla_{\rho}n'^{\nu}\right)\nonumber\\
 &=& -\frac{1}{\epsilon\gamma}
 n'^{\mu}\left(M^{\nu}n'^{\rho}\nabla_{\rho}n'_{\nu}+
   n'_{\nu}M^{\rho}\nabla_{\rho} n'^{\nu}\right)\nonumber\\
&=& -\frac{1}{\epsilon\gamma} n'^{\mu}
\left(2M^{\nu}n'^{\rho}\nabla_{\rho}n'_{\nu}+ n'^{\nu}M^{\rho} ({\rm
    d}n')_{\rho\nu}\right)\nonumber\\
&=& -\frac{1}{\epsilon\gamma} n'^{\mu}\left(2M^{\nu}\sqrt{|\gamma|}
  \tilde{n}'^{\rho}\nabla_{\rho}(\sqrt{|\gamma|}\tilde{n}'_{\nu})+
  2n'^{\nu}M^{\rho}n'_{[\rho}w_{\nu]}\right)\nonumber\\
&=& \frac{1}{\epsilon\gamma} n'^{\mu}n'^{\nu}n'_{\nu}M^{\rho}w_{\rho}=
n'^{\mu}M^{\rho}w_{\rho}  \label{nXnormal}
\end{eqnarray}
using $M^{\nu}\tilde{n}'_{\nu}=0$ and the geodesic property
$\tilde{n}'^{\rho}\nabla_{\rho}\tilde{n}'_{\nu}=0$ of the normal in a Gaussian
system. With
\begin{equation}
 v^{\rho}({\rm d}n')_{\rho\nu}= 2(N n'^{\rho}+M'^{\rho}) n'_{[\rho} w_{\nu]}=
\epsilon \gamma N w_{\nu}- M'^{\rho}w_{\rho} n'_{\nu}\,,
\end{equation}
we can write (\ref{gGaussprime}) as
\begin{equation}
 0= n'^{\mu}g_{\mu\nu}
 n'^{\rho}\partial_{\rho}N + [n',M']^{\mu}g_{\mu\nu}+\partial_{\nu}(N
 n'^{\mu}n'^{\rho}g_{\mu\rho})+  \epsilon\gamma N
 w_{\nu}- M'^{\rho}w_{\rho} n'_{\nu}
\end{equation}
or
\begin{equation}
 0=[n',M']^{\mu}+n'^{\mu}n'^{\rho}\partial_{\rho}N
 +\epsilon\partial^{\mu}(\gamma N)
+ \epsilon\gamma N w^{\mu}- M'^{\rho}w_{\rho} n'^{\mu}\,.
\end{equation}

The equation can now be split into components parallel and orthogonal to
$n'^{\mu}$: The normal component implies
\begin{equation} \label{nrho}
  n'^{\rho}\partial_{\rho}N = -\frac{1}{2}\frac{N}{\gamma}
  n'^{\nu}\partial_{\nu}\gamma
\end{equation}
(the contribution from ${\rm d}n'$ cancelling out with the normal contribution
from $[n',M']$) while the spatial component gives
\begin{equation} \label{nXa}
[n',M']^a = q^a_{\mu}[n',M']^{\mu}= - \epsilon q^{ab}\partial_b (\gamma N) -
\epsilon\gamma N w^a =-\epsilon ({\rm
  grad}_q(\gamma N))^a - \epsilon\gamma N w^a\,.
\end{equation}
The full space-time commutator is
\begin{equation} \label{nX}
 [n',M']^{\mu}=  q^{\mu}_a [n',M']^a + M'^{\rho}w_{\rho}n'^{\mu}\,,
\end{equation}
combining (\ref{nXa}) with (\ref{nXnormal}).

With these relations, the hypersurface-deformation brackets follow immediately
from the Lie brackets of $g$-Gaussian vector fields. First, in the Gaussian
system, (\ref{nrho}) and (\ref{nX}) provide first-order partial differential
equations for $N$ and $M^{\mu}$ or $M'^{\mu}$ to be extended into a
neighborhood of the initial slice. (Importantly, all $M^{\mu}$-dependent terms
cancel out in (\ref{nrho}) even with non-standard normalization. The equation
for $N$ is therefore decoupled from the equation for $M^{\mu}$.) We can
then compute space-time Lie brackets of two $g$-Gaussian vector fields
\begin{eqnarray}
  [v_1,v_2] &=& [N_1n'+M'_1,N_2n'+M'_2]\nonumber\\
  &=& [N_1n',N_2n']+[N_1n',M'_2]+[M'_1,N_2n']+ [M'_1,M'_2]\\
  &=& (N_1{\cal L}_{n'}N_2- N_2{\cal L}_{n'}N_1)n'+ ({\cal
    L}_{M'_1}N_2- {\cal L}_{M'_2}N_1)n'+
  N_1[n',M'_2]-N_2[n',M'_1]+ [M'_1,M'_2]\,.\nonumber
\end{eqnarray}
The first term, $N_1{\cal L}_{n'}N_2- N_2{\cal L}_{n'}N_1$, is zero even with
the new contributions in (\ref{nrho}) for non-constant $\gamma$. Similarly,
the $w^a$-term in (\ref{nXa}) does not contribute to
$N_1[n',M'_2]-N_2[n',M'_1]$. However, the normal contribution
$M'^{\rho}w_{\rho}n'^{\mu}= -\frac{1}{2}\gamma^{-1}
n'^{\mu}M'^{\rho}\partial_{\rho}\gamma $ in (\ref{nX}) does not cancel out and
provides a new normal term in
\begin{eqnarray}
[v_1,v_2]  &=& \left({\cal L}_{M'_1}N_2-{\cal
      L}_{M'_2}N_1-\frac{1}{2\gamma}\left(N_1{\cal L}_{M'_2}\gamma-
      N_2{\cal
      L}_{M'_1}\gamma \right)\right)n'\nonumber\\
 &&- \epsilon N_1
  {\rm grad}_q(N_2\gamma) +  \epsilon N_2{\rm grad}_q(N\gamma)
 +  [M'_1,M'_2]\nonumber\\
&=&\frac{1}{\sqrt{|\gamma|}} \left({\cal L}_{M'_1}(\sqrt{|\gamma|}N_2)-
{\cal L}_{M'_2}(\sqrt{|\gamma|}N_1)\right)n'\nonumber\\
&&-\epsilon \gamma (N_1{\rm
  grad}_qN_2-N_2{\rm
  grad}_qN_1) +[M'_1,M'_2]\,. \label{vv}
\end{eqnarray}

The last line can now be transformed from $n'^{\mu}=n^{\mu}-\alpha^{\mu}$ and
$M'^{\mu}=M^{\mu}+N\alpha^{\mu}$ to $n^{\mu}$ and $M^{\mu}$. Inserting the
expressions for the primed vectors leads to several extra terms, most of which
cancel out. However, two new contributions remain:
\begin{eqnarray}
[v_1,v_2]
&=&\left(\frac{1}{\sqrt{|\gamma|}} \left({\cal L}_{M_1}(\sqrt{|\gamma|}N_2)-
{\cal L}_{M_2}(\sqrt{|\gamma|}N_1)\right)+ N_1{\cal L}_{\alpha}N_2- N_2{\cal
L}_{\alpha}N_1 \right)n \label{vv2}\\
&&-\epsilon\gamma (N_1{\rm  grad}_qN_2-N_2{\rm grad}_qN_1) -\sqrt{|\gamma|}
\left(N_1{\cal L}_{M_2}\frac{\alpha}{\sqrt{|\gamma|}}- N_2 {\cal
    L}_{M_1}\frac{\alpha}{\sqrt{|\gamma|}}\right) + [M_1,M_2]\,.\nonumber
\end{eqnarray}
By extracting terms parallel to $n$ or the tangent plane, we write this Lie
bracket as bracket relationships between pairs $(N,M^a)$:
\begin{eqnarray}
 [(0,M_1^a),(0,M_2^b)] &=& (0,[M_1,M_2]^c)\\\mbox{}
 [(N,0),(0,M^a)] &=& \left(-|\gamma^{-1/2}| {\cal L}_{M} (|\gamma|^{1/2}N),
   -|\gamma|^{1/2} N {\cal L}_M (|\gamma|^{-1/2}\alpha^a)\right)\label{NM}\\
\mbox{}
 [(N_1,0),(N_2,0)] &=& \left(N_1{\cal L}_{\alpha}N_2-N_2{\cal L}_{\alpha} N_1,
   -\epsilon\gamma(N_1{\rm grad}^a_qN_2-N_2{\rm
     grad}^a_qN_1)\right)\,. \label{NN}
\end{eqnarray}

The following special cases are of interest:
\begin{itemize}
\item If $\alpha^a\not=0$, there is a new class of modified brackets which
  have not been derived explicitly in models of loop quantum gravity. New
  features are a transversal deformation (along a non-normal $n^{\mu}$)
  contributing to the bracket of two transversal deformations, and a spatial
  diffeomorphism contributing to the bracket of a transversal deformation and
  a spatial diffeomorphism. If this example is realized by quantum-gravity
  effects, it would require the existence of a preferred spatial direction
  $\alpha^a$.
\item If $\alpha^a=0$, the bracket of two normal deformations is a spatial
  diffeomorphism, as in the classical version, but with a multiplicative
  correction function $\gamma=\beta$. One can obtain the modified brackets
  (\ref{NN}) by replacing $N_i$ with $\sqrt{|\gamma|}N_i$ and $n'$ with
  $n'/\sqrt{|\gamma|}$ in the standard brackets, in accordance with the
  rescaling transformations of the normal keeping $N n'$ invariant for
  (\ref{eq:taudecompmod}) to be preserved. However, our calculation shows more
  than this because it ensures that the three conditions required for a
  meaningful relation between hypersurface-deformation brackets and space-time
  Lie brackets are still satisfied for $g$-Gaussian vector fields with a
  non-standard normal.
\item If $\alpha^a=0$ and $\gamma=\beta$ is spatially constant, all
  derivatives of $\gamma$ cancel out and the bracket of a normal deformation
  and a spatial diffeomorphism is unmodified.  A time-dependent $\gamma$
  therefore leads only to a multiplicative modification of the standard
  brackets, and it appears only in the bracket of two normal
  deformations. This is the example (\ref{HHbeta}) found in models of loop
  quantum cosmology with first-order perturbative inhomogeneity. 
\end{itemize}

\subsection{Lie algebroids}
\label{s:LieAlg}

The hypersurface-deformation generators do not form a Lie algebra, owing to
the appearance of structure functions. Structure functions can be elegantly
described by the notion of Lie algebroids, which may be motivated as follows:
Assume that we have a finite number of constraints $C_I$, $I=1,\ldots,n$, on a
Poisson manifold $B$, which satisfy an algebra $\{C_I,C_J\}=c_{IJ}^K(x)C_K$
with structure functions $c_{IJ}^K(x)$ depending on $x\in B$. We can formally
rewrite brackets with structure functions in terms of structure constants by
defining an extended system of infinitely many constraints
\begin{eqnarray} \label{CIJH}
 C_I\quad,\quad&& C_{IJ}:= \{C_I,C_J\}= c_{IJ}^KC_K\quad,\quad \nonumber\\
&&C_{HIJ}:=
 \{C_H,C_{IJ}\}=(\{C_H,c_{IJ}^K\}+c_{IJ}^Lc_{HL}^K)C_K \quad, \quad \ldots
\end{eqnarray}
The brackets $\{C_I,C_J\}=C_{IJ}$, $\{C_H,C_{IJ}\}=C_{HIJ}$, \ldots\ of the
extended system then have structure constants.

These constraints span a certain linear subspace of the space $\Gamma(A)$ of
sections $\alpha=\alpha(x)^IC_I$ of a vector bundle $A$ over the base manifold
$B$ (phase space) with fiber $\pi^{-1}(x)\approx{\mathbb R}^n\ni
\{\alpha(x)^1,\ldots,\alpha(x)^n\}$. The sections of this bundle form a Lie
algebra by taking Poisson brackets
$[\alpha_1,\alpha_2]=\{\alpha_1(x)^IC_I,\alpha_2(x)^JC_J\}$. Moreover, we can
define a linear map $\rho\colon\Gamma(A)\to \Gamma(TB),
\alpha=\alpha^I(x)C_I\mapsto \{\alpha(x)^IC_I,\cdot\}$ which appears in a
Leibniz rule
\begin{eqnarray}
 [\alpha,g\beta] &=&\{\alpha(x)^IC_I,g(x)\beta(x)^JC_J\} =
g(x) \{\alpha(x)^IC_I,\beta(x)^JC_J\}+
 \{\alpha(x)^IC_I,g(x)\} \beta(x)^JC_J\nonumber\\
 &=& g(x) \{\alpha(x)^IC_I,\beta(x)^JC_J\}+
 \left(\rho(\alpha(x)^IC_I)g(x)\right) \beta(x)^JC_J\nonumber\\
 &=& g[\alpha,\beta]+ (\rho(\alpha)g)\beta\,,
\end{eqnarray}
and $\rho$ is a homomorphism of Lie algebras:
\begin{eqnarray}
 \rho([\alpha,\beta]) &=& \{\{\alpha(x)^IC_I,\beta(x)^JC_J\},\cdot\}\nonumber\\
 &=& \{\alpha(x)^IC_I,\{\beta(x)^JC_J,\cdot\}\}-
 \{\beta(x)^JC_J,\{\alpha(x)^IC_I,\cdot\}\}\nonumber\\
 &=&
 \rho(\alpha)\rho(\beta)-\rho(\beta)\rho(\alpha)=[\rho(\alpha),\rho(\beta)]
\end{eqnarray}
using the Jacobi identity.  The Lie bracket on sections together with a
homomorphism $\rho$ characterize $A$ as a Lie algebroid \cite{Pradines}.

\begin{defi}
  A {\em Lie algebroid} is a vector bundle $A$ over a smooth base manifold $B$
  together with a Lie bracket $[\cdot,\cdot]_{A}$ on the set $\Gamma(A)$ of
  sections of $A$ and a bundle map $\rho\colon \Gamma(A)\rightarrow
  \Gamma(TB)$, called the {\em anchor}, provided that
\begin{itemize}
\item $\rho\colon\left(\Gamma(A),[\cdot,\cdot]_{A}\right)\rightarrow
  \left(\Gamma(TB),[\cdot,\cdot]\right)$ is a homomorphism
  of Lie algebras, that is
\[
\rho\left(\left[\xi,\eta\right]_{A}\right) =
\left[\rho(\xi),\rho(\eta)\right]\,,
\]
where $[\cdot,\cdot]$ is the commutator of vector fields in $\Gamma(TB)$.
\item For any $\xi$, $\eta\in\Gamma(A)$ and for any $f\in C^{\infty}(B)$ the
  Leibniz identity
\[
\left[\xi,f\eta\right]_{A}=f[\xi,\eta]_{A}+\left(\rho(\xi)
  f\right)\eta
\]
holds.
\end{itemize}
\end{defi}

If the base manifold $B$ is a point, the Lie algebroid is a Lie
algebra. Another example for a Lie algebroid is the tangent bundle $TB$ of a
manifold $B$ with $\rho\colon \Gamma(TB)\rightarrow \Gamma(TB)$ the identity
map and the Lie bracket of vector fields as the bracket on sections. The
hypersurface-deformation brackets have been shown in \cite{ConsAlgebroid} to
be captured by a certain Lie algebroid more specific than the construction
based on (\ref{CIJH}). This notion can therefore provide useful methods in an
analysis of different versions of hypersurface deformations.  In order to
identify classes of equivalent Lie algebroids, one may generalize the notion
of a Lie algebra morphism to the Lie algebroid case.
\begin{defi}
  A {\em base-preserving morphism} between Lie algebroids
  $\left(A,[\cdot,\cdot]_{A},\rho\right)$ and
  $\left(A',[\cdot,\cdot]_{A'},\rho'\right)$ is a bundle map $\Phi\colon
  A\rightarrow A'$ over ${\rm id}_B\colon B\rightarrow B'=B$ such that $\Phi$
  induces a Lie algebra homomorphism
  $\Phi\colon\left(\Gamma(A),[\cdot,\cdot]_{A}\right)\rightarrow
  \left(\Gamma(A'),[\cdot,\cdot]_{A'}\right)$ and satisfies $\rho'\circ\Phi =
  \rho$.
\end{defi}

If the induced base map $\phi_0$ is a diffeomorphism, the definition can still
be used. In such cases, which will be of interest to us, the bundle map
induces a map on sections via $\Phi(\xi)(y)= \xi(\phi_0^{-1}(y))$ for
$\xi\in\Gamma(A)$ and $y\in B'$.  For completeness, we mention that a Lie
algebroid morphism which does not preserve the base manifold can be defined as
follows; see for instance \cite{LieAlgMorph}:

\begin{defi}
  A {\em Lie algebroid morphism} from $A\to B$ to $A'\to B'$ is a
  bundle map $\phi\colon A'^*\to A^*$ with induced base map $\phi_0\colon
  B'\to B$, such that:
\begin{enumerate}
 \item The induced map $\Phi\colon \Gamma(A)\to\Gamma(A')$, defined by
   $\Phi(\xi)(y)= \phi^* \xi(\phi_0(y))$ for $y\in B'$, preserves the
   Lie bracket on sections: $[\Phi(\xi),\Phi(\eta)] =
   \Phi([\xi,\eta])$ for all $\xi,\eta\in \Gamma(A)$.
\item We have $\rho= \phi_{0*}\circ \rho'\circ\Phi$.
\end{enumerate}
\end{defi}

We will not use general morphisms in this paper, but note that an example of a
morphism as in the preceding definition could be used to relate the space-time
structures underlying general relativity and higher-curvature actions,
respectively. The latter are higher-derivative theories and have additional
canonical degrees of freedom compared with general relativity; therefore, the
base manifolds are not diffeomorphic. Nevertheless, the
hypersurface-deformation brackets are the same in both settings
\cite{HigherCurvHam} and could be used to construct a Lie algebroid morphism.

From now on, we focus on the specific example of the algebroid underlying
general relativity. We quote useful definitions and one central result from
\cite{ConsAlgebroid}:
\begin{itemize}
\item A connected Lorentzian manifold (or space-time) $(\mathcal{M},g)$ is
  called $\Sigma$-{\em adapted} if it admits an embedding of $\Sigma$ as a
  spacelike hypersurface. Such an embedding is called a $\Sigma$-{\em space}
  in $\mathcal{M}$, and a pair consisting of a space-time and a $\Sigma$-space
  in it is called a $\Sigma$-{\em space-time}. On every $\Sigma$-space we have
  an induced, or spatial, metric $q=i^*g$ using the embedding $i\colon
  (\Sigma,q)\hookrightarrow ({\cal M},g)$.
\item Coordinate independence leads to the concept of a {\em
    $\Sigma$-universe}, an equivalence class $[i]$ of $\Sigma$-space-times
  where
  $i\colon\left(\Sigma,q\right)\hookrightarrow\left(\mathcal{M},g\right)$ and
  $i'\colon\left(\Sigma,q\right)\hookrightarrow\left(\mathcal{M}',g'\right)$
  are equivalent if there is an isometry
  $\Psi\colon\left(\mathcal{M},g\right)\rightarrow\left(\mathcal{M}',g'\right)$
  which preserves the coorientation of $\Sigma$ and satisfies $\Psi\circ
  i=i'$. The set of all $\Sigma$-universes is denoted by $\mathcal{U}\Sigma$.
  In order to confirm that this definition is consistent, we pull back $g'$
  along $i'$ and obtain the same result as before applying the isometry:
  $\left(i'\right)^{*}g' = \left(\Psi\circ i\right)^{*}g' =
  i^{*}\left(\Psi^{*}g'\right)=i^*g = q$.
\item So far, the relations between a Cauchy hypersurface $\Sigma$ and
  space-time $\mathcal{M}$ have been formalized. The next step is to look at
  the evolutions of one time slice into another time slice. A time slice is
  defined to be an embedding $i_t$ for a fixed time parameter $t={\rm
    constant}$ within a 1-parameter family. Different time slices are related
  by {\em $\Sigma$-evolutions}, equivalence classes $[i_1,i_0]$ of pairs
  $(i_1,i_0)$ of $\Sigma$-spaces in the same space-time, where a pair
  $(i_1,i_0)$ in $\mathcal{M}$ is equivalent to $(i'_1,i'_0)$ in
  $\mathcal{M}'$ if there is a single isometry $\Psi\colon
  \mathcal{M}\rightarrow \mathcal{M}'$ which is consistent with the
  coorientations of time slices and which satisfies both $\Psi\circ i_1 =
  i'_1$ and $\Psi\circ i_0=i'_0$. The set of all $\Sigma$-evolutions is
  denoted by $\mathcal{E}\Sigma$.
\end{itemize}

The set of $\Sigma$-evolutions, $\mathcal{E}\Sigma$, forms a Lie groupoid
\cite{ConsAlgebroid} with elements in $\mathcal{U}\Sigma$, source map
$s([i_1,i_0])=[i_0]$ and target map $t([i_1,i_0])=[i_1]$, multiplication given
by $[i_2,i_1][i_1,i_0]=[i_2,i_0]$ and inversion by
$[i_1,i_0]^{-1}=[i_0,i_1]$. The definition therefore gives rise to an
evolution picture in terms of groupoid multiplication.  The Lie algebroid
$A\mathcal{E}\Sigma$ belonging to the Lie groupoid $\mathcal{E}\Sigma$
provides the link between this formulation and the infinitesimal one used for
instance in \cite{Regained}.  According to \cite{ConsAlgebroid},

\begin{prop}
  The Lie algebroid $A\mathcal{E}\Sigma$ of ${\cal E}\Sigma$ is isomorphic as
  a vector bundle to the trivial bundle
  $\mathcal{U}\Sigma\times\left(\Gamma(T\Sigma)\oplus
    C^{\infty}(\Sigma)\right)$ over the base manifold $\mathcal{U}\Sigma$.
\label{LieAlg}
\end{prop}

Proposition~\ref{LieAlg} tells us that infinitesimal evolutions of an
equivalence class in $\mathcal{U}\Sigma$ are described by (shift) vector
fields in $\Gamma(T\Sigma)$ and (lapse) $C^{\infty}$-functions on
$\Sigma$. The base manifold of the Lie algebroid is the space of equivalence
classes of spatial embeddings. Structure functions of the classical
hypersurface-deformation brackets depend on the spatial metric, which in turn
depends only on the equivalence class of embeddings
$\Sigma\hookrightarrow{\cal M}$ for a given space-time metric. Similarly,
extrinsic curvature on $\Sigma$ depends on the embedding in $({\cal M},g)$,
but it is not invariant under space-time isometries fixing $(\Sigma,q)$. Since
the modification function $\beta$ may depend on all phase-space variables, we
should refine the equivalence classes to those transformations that keep both
$q_{ab}$ and $K_{cd}$ fixed on $\Sigma$.  However, if the
hypersurface-deformation brackets are modified, it is not clear whether a
space-time metric structure exists which can induce a spatial metric. It is
then more appropriate to formulate the Lie algebroid directly over a base
manifold of spatial metrics and extrinsic-curvature tensors on $\Sigma$ (or
the classical phase space). In fact, \cite{ConsAlgebroid} indicates the way to
such a formulation using Gaussian representatives.

For an explicit construction of Lie algebroid brackets and the anchor,
\cite{ConsAlgebroid} chooses as a representative for a $\Sigma$-universe a
slicing which is locally of Gaussian form, as in the derivation of
Sec.~\ref{s:brief}. A representative of a class in ${\cal U}\Sigma$ can then
be fixed by specifying the induced metric $q$ instead of the embedding. The
tangent space of the resulting base manifold of spatial metrics is, at a point
$q$, given by $T_q{\cal U}\Sigma= S^2T^*\Sigma$, the space of symmetric
tensors identified with Lie derivatives of the space-time metric by
$g$-Gaussian vector fields $v^{\mu}=N n^{\mu}+M^{\mu}$: Since such vector
fields preserve the Gaussian form, ${\cal L}_vg$ is equivalent to a change
$\delta_vq:={\cal L}_Mq+N\dot{q}$ of just the spatial metric, where
$\dot{q}={\cal L}_nq=2K$ is proportional to the extrinsic-curvature tensor.
The latter changes by $\delta_vK={\cal L}_MK+N\dot{K}(q,K)$ where
$\dot{K}={\cal L}_nK$ is a function of $q_{ab}$ and $K_{cd}$ via the field
equations. (The field equations had been bypassed on \cite{ConsAlgebroid} by
working with equivalence classes of entire neighborhood of embeddings of
$\Sigma$ in $M$.) Notice that the anchor $\rho$ depends on the field equations
of the theory, while the brackets do not.

The anchor map of the Lie algebroid with the gravitational phase space as base
manifold is given by $(N,M)\mapsto (\delta_{N n+M}q,\delta_{Nn+M}K)$. This
base manifold and anchor have been extended to the space of induced metrics
and extrinsic-curvature tensors, which is necessary if one works with modified
brackets where $\beta$ depends on $q_{ab}$ and $K_{ab}$.  The same
calculations as in Sec.~\ref{s:brief} imply that the Lie algebra of
$g$-Gaussian vector fields $v$ leads to a Lie-algebroid bracket
\begin{eqnarray}
&&[(N_1,M_1),(N_2,M_2)] \\
&=&\left(\frac{1}{\sqrt{|\beta|}} \left({\cal
    L}_{M_1}(\sqrt{|\beta|}N_2)-
{\cal L}_{M_2}(\sqrt{|\beta|}N_1)\right),
\epsilon\beta (N_1{\rm
  grad}_qN_2-N_2{\rm
  grad}_qN_1) +[M_1,M_2]\right)\nonumber
\end{eqnarray}
(if $\alpha^a=0$) once the decomposition $v^{\mu}=N n^{\mu}+M^{\mu}$ is
introduced.

\section{Physics from hypersurface-deformation algebroids}
\label{s:Def}

Using the Lie-algebroid structure of hypersurface deformations, we can now
look at possible modified versions and their relations to the classical
brackets. In some cases, they turn out to be related by algebroid
morphisms. We begin with a review of existing examples for deformed brackets.

\subsection{Modified brackets}

The classical hypersurface-deformation brackets have been derived from the
usual space-time structure, using for instance infinitesimal space-time
diffeomorphisms in (\ref{vv}). They are independent of specific solutions to
Einstein's or modified field equations as long as the theory is based on
Riemannian geometry. For instance, the same brackets are obtained for
higher-curvature actions \cite{HigherCurvHam}. In several effective models of
loop quantum gravity, however, modified versions of the brackets have been
found, and it has not been clear what space-time structure or what effective
actions they may correspond to. In this subsection, we discuss several
relevant conceptual details of such models, leaving aside technical features.

Modified brackets have been derived canonically, by including possible quantum
corrections in the classical constraints and checking under which conditions
they still give rise to a closed set of Poisson brackets. Generically, quantum
corrections suggested by loop quantum gravity, based on real connection
variables, could be implemented consistently only when the brackets were
modified as in (\ref{HHbeta}). For complex connections, the derivative
structure of the Hamiltonian constraint is different, in that there are no
second-order derivatives of the triad unlike in real formulations which have
the generic pattern responsible for signature-change type deformations
\cite{Loss}. At least in spherically symmetric models, it is then possible to
have undeformed brackets even in the presence of holonomy modifications
\cite{SphSymmComplex}. Such models are less restrictive than the full theory,
and therefore it is not clear whether the full brackets can be undeformed.

Two main classes of models in which deformed brackets have been derived are:
(i) cosmological perturbations \cite{ConstraintAlgebra,ScalarHolInv} where, to
linear order, $\beta$ is a function only of time (via the background spatial
metric and extrinsic curvature) and (ii) spherically symmetric models
\cite{JR,LTBII,ModCollapse,HigherSpatial} where $\beta$ may also depend on the
radial coordinate. With so-called holonomy modifications of the classical
dynamics, $\beta$ depends on $K_{ab}$ as some kind of higher-curvature
correction, but only in spatial terms so that the modification is not
necessarily space-time covariant. Detailed calculations have shown that it is
possible to have such spatial-curvature modifications and still maintain
closed brackets of correspondingly modified hypersurface-deformation
generators, but only when $\beta$ and the way it appears in the equations of
motion are restricted. This is the condition of anomaly-freedom.  Generically,
whenever $\beta$ depends on $K_{ab}$, it changes sign at large curvature if
quantum effects lead to bounded curvature or densities (so-called bounce
models). The same observations have been found in cosmological and spherically
symmetric models, with agreement also in the specific functional form of
$\beta$ \cite{DeformedCosmo}. There are, however, obstructions in models with
local physical degrees of freedom \cite{SphSymmCov,GowdyCov}, in which no
anomaly-free holonomy-modified versions have been found yet. (There are also
obstructions in some operator versions of spherically symmetric models that
implement spatial discreteness \cite{SphSymmDisc}.)

In these two classes of models, two kinds of methods have been used to provide
complementary insights: Effective calculations proceed by computing Poisson
brackets of classical hypersurface-deformation generators modified by
potential quantum corrections, following a systematic canonical version of
effective-action techniques \cite{EffAc,HigherTime,EffConsQBR,CW}. Operator
methods compute commutators of quantized generators. Also here, there is full
agreement between results from these two different methods: The operator
calculations of \cite{SphSymmOp} in spherically symmetric models provide the
same restrictions on modifications and the function $\beta$ as found by
effective methods \cite{JR}. It is not known how to implement cosmological
perturbations at the operator level, but there is a set of $2+1$-dimensional
models which provide complementary insights. In \cite{ThreeDeform}, a
modification function for holonomies has been found that shows the same
features related to the change of sign of $\beta$; see also
\cite{TwoPlusOneSymm}.

Other operator calculations in $2+1$-dimensional models
\cite{TwoPlusOneDef,TwoPlusOneDef2,AnoFreeWeak} are only partially off-shell
so far, and therefore are not able to show the full brackets. In particular,
since they amount to factoring out spatial diffeomorphisms everywhere except
at a finite number of isolated points, they cannot exhibit holonomy
modifications which are spatially non-local. The interesting conclusion of
$\beta$ changing sign therefore cannot yet be tested in this
setting. Nevertheless, these models have confirmed the presence of modified
brackets for metric-dependent modifications. For instance, Eq.~(9.27) in
\cite{TwoPlusOneDef} gives a definition of the right-hand side of the operator
equivalent of (\ref{HHbeta}), which contains an inverse-metric operator with a
factor of $(\det q)^{-1/4}$ modified by so-called inverse-triad corrections
\cite{QSDI,InvScale}.  We note that reading off modified brackets from
commutators is not straightforward because in addition to the commutator, an
effective bracket contains information about semiclassical states. Defining
such states and computing expectation values in them is notoriously difficult
in background-independent quantum-gravity theories. Nevertheless, it is clear
that the naive classical limit of the equation just cited shows a modification
of the classical bracket. (In the naive classical limit, one replaces operator
factors in the quantized constraints and structure functions with their
expectation values in simple states, thus ignoring fluctuations and higher
moments.)

Some quantization schemes of constrained gravitational systems represent
hypersurface deformations in an indirect way, after reformulating the
classical constraints so as to make them easier to quantize. In the present
context, two examples are relevant in which one can use reformulations in
order to eliminate structure functions from the constraint brackets. In
\cite{AnoFreeWeak}, $2+1$-dimensional gravity is quantized by writing the
bracket of two Hamiltonian constraints in the schematic form
$\{H[N],H[M]\}=\{D[N'^a],D[M'^b]\}$ where $N'^a$ and $M'^b$ are shift vector
fields related to $N$ and $M$, respectively. There are no structure functions
on the right-hand side, and it is possible to represent the bracket relation
without modifications. However, this result does not imply that the
hypersurface-deformation brackets are undeformed; in fact, one can check that
$\{H[N],H[M]\}$ written as a single diffeomorphism constraint has
quantum-corrected structure functions. (The vector fields $N'^a$ and $M'^b$
mentioned above depend on the spatial metric and give rise to new terms in
structure functions when $\{D[N'^a],D[M'^b]\}$ is expressed as a term linear
in $D$.)

Similarly, spherically symmetric systems can be reformulated in a way that
partially Abelianizes the constraint algebra \cite{LoopSchwarz,LoopReiss}. The
Hamiltonian constraint is here replaced by a linear combination
$C[L]:=H[L']+D[L'']$ with $L'$ and $L''$ suitably related to $L$ such that
$\{C[L_1],C[L_2]\}=0$. Structure functions are thus eliminated from the
constrained system $(C,D)$, and the brackets can be represented without
quantum corrections in their coefficients. However, if one tries to find
hypersurface-deformation generators of quantum constraints with the correct
classical limit, it turns out that this is possible only if the
hypersurface-deformation brackets are deformed \cite{SphSymmCov,GowdyCov}.

Since all these examples are obtained after quantizing generators of normal
deformations with respect to $n^{\mu}$ such that
$g_{\mu\nu}n^{\mu}n^{\nu}=\epsilon$ and the vector field $n^{\mu}$ is not
subject to quantum corrections, the deformed algebra refers to a unit normal
vector. With such modified brackets but standard normalization, the space-time
considerations of \cite{Regained} no longer apply, and therefore a
non-classical space-time structure seems to be realized.

The new brackets, in general, cannot be viewed as describing deformations of
hypersurfaces in a Riemannian space-time with metric $g_{\mu\nu}$. They
do, however, determine a well-defined canonical theory, in which one can, in
principle, solve the constraints and compute gauge-invariant observables,
which is all that is needed for physical predictions. Importantly, the
brackets are still closed, which is the challenging part of their
constructions. If the brackets were not closed, the models would be anomalous
and inconsistent because gauge transformations would be violated and results
would depend on choices of coordinates.

Modified brackets can be formulated as a Lie algebroid over the space of pairs
of symmetric tensor fields $(q_{ab},K_{cd})$ with positive-definite $q_{ab}$.
The inverse of $q_{ab}$, as well as $K_{ab}$ through possible modifications in
$\beta$, appear in the structure functions of the constraint brackets, but
they play the role only of phase-space functions which need not have a
geometrical interpretation as spatial metric and extrinsic curvature
associated with a slice $\Sigma$ in space-time $({\cal M},g)$.  Instead of
defining these spatial tensors in terms of the embedding functions $X(x)$ and
a space-time metric $g_{\mu\nu}$, the only option is to view $q_{ab}$ and
$K_{ab}$ as independent phase-space degrees of freedom on which the
constraints and the structure functions depend. The modification function must
be covariant under transformations with brackets (\ref{DD}), (\ref{HD}) and
(\ref{HHbeta}). In particular, since these brackets contain infinitesimal
spatial diffeomorphisms as a subalgebra, $\beta$ must be a spatial scalar. In
the modified case, the theory is not necessarily standard space-time
covariant, but if the brackets close, $\beta$ and the resulting theory are
covariant under transformations generated by Poisson brackets with the
modified constraints. In the absence of a space-time picture, the physical
meaning of $q_{ab}$ and $K_{cd}$ is supplied by how they appear in canonical
observables. The latter have a known interpretation in the classical limit of
$\beta\to 1$ (low curvature), which is extended to non-classical regimes in an
anomaly-free deformed theory. Alternatively, one may employ field
redefinitions such that a relation of Lie algebroid elements to space-time
metrics becomes possible. We discuss two possible types in the following
subsections.

\subsection{Base transformations}

In (\ref{HHbeta}), $\beta$ always appears in combination with the inverse of
$q^{ab}$, whose components can be used as coordinates on the base manifold
along with the components of $K_{ab}$. We can define a transformation of the
base manifold by mapping $(q_{ab},K_{cd})$ to $(|\beta|^{-1}q_{ab},K_{cd})$
and extend it to a fiber map $(q_{ab},K_{cd},N,M^e)\mapsto
(|\beta|^{-1}q_{ab},K_{cd},N,M^e)$. Here, the fiber coordinates $N$ and
$M^e$ as well as $K_{cd}$ are unchanged, while $q_{ab}$ absorbs $|\beta|$. As
long as $\beta\not=0$, the base map is a diffeomorphism and a well-defined Lie
algebroid morphism is obtained, eliminating $|\beta|$ from the brackets. The
only parameter that cannot be absorbed is ${\rm sgn}\beta$ because $q_{ab}$ is
required to be positive definite and, in particular, invertible.

We may then consider $|\beta|^{-1}q_{ab}$ as the spatial metric on a spatial
slice in a space-time with line element
\begin{equation} \label{dsbeta}
 {\rm d}s^2=\epsilon\epsilon_{\beta} N^2{\rm d}t^2+ |\beta|^{-1} q_{ab}
 ({\rm d}x^a+M^a{\rm d}t)({\rm d}x^b+M^b{\rm d}t)
\end{equation}
which generically cannot be obtained by a coordinate transformation from
(\ref{ds}). (If this were possible, one could eliminate the scale factor
$a=|\beta|^{-1/2}$ of a Friedmann--Robertson--Walker metric by a coordinate
transformation.)  The extrinsic curvature of a $t={\rm constant}$ slice in
(\ref{dsbeta}) is not equal to $K_{ab}$. However, we can use the field
equations of the modified theory in order to relate $K_{ab}$ to
$\dot{q}_{ab}={\cal L}_nq_{ab}$. Using the standard equation for extrinsic
curvature computed from (\ref{dsbeta}), a relationship between $K_{ab}$ and
extrinsic curvature is obtained, which may not be the identity.

The new variables $(|\beta|^{-1}q_{ab},K_{cd})$ are no longer canonical
coordinates on the base manifold. Non-canonical base coordinates do not make a
difference for a Lie algebroid, which in general does not even have a Poisson
structure on its base. However, we need a Poisson structure on the base
manifold in order to derive the dynamics generated by the constraints, and for
this it is useful to have a canonical set of variables.  Modifying the map
$(q_{ab},K_{cd}) \mapsto (|\beta|^{-1}q_{ab},K_{cd})$ such that it becomes
canonical is possible in some models \cite{Absorb}, but may be complicated in
general.

While base transformations can map modified brackets to the classical version,
as long as $\beta$ does not change sign, it is not easy to derive general,
theory-independent effects because the interpretation of $K_{ab}$ depends on
the dynamics, and there may be no simple canonical sets of variables.  It
turns out that general aspects of physical implications of the absorption are
easier to discern if one uses morphisms that originate from fiber maps. We
will be able to do so by absorbing $|\beta|$ in the normalization condition,
at least partially, allowing us to discuss possible physical implications in
general terms.

\subsection{Change of normalization as algebroid morphism}

One usually expects that the classical theory can be recovered when $\beta$
approaches one in some regime, such as low curvature.  However, as already
mentioned, the classical theory can be described with a more general $\beta$
if one uses non-standard normalizations $g_{\mu\nu}n^{\mu}n^{\nu}=
\epsilon\beta$ of normal vectors to hypersurfaces.  Even the classical
brackets can therefore be modified without changing the implied physics.
Although it is customary to assume the normal vector $n^{\mu}$ to be
normalized to $\epsilon=\pm 1$, depending on the signature, this choice is a
mere convention and one may as well introduce a different normalization.
Thus, the requirement of having the correct classical limit does not restrict
$\beta$ much, except that $\beta$ should not be identically zero.

Since we know from Sec.~\ref{s:brief} that, for spatially constant $\beta$,
the hypersurface-deformation brackets belong to a Lie algebroid, irrespective
of how the normal is normalized, there are no further conditions on $\beta$
from the Jacobi identity. As in our explicit derivation of the brackets, we
may obtain a deformation by using a non-standard normalization of the normal
vector field in classical general relativity.

We introduce a bundle map $\Phi$ with fiber map $(N,M^a)\mapsto
(\sqrt{|\beta|}N,M^a)$ and the identity as base map. It obeys
\begin{eqnarray}
[\Phi((N_1,0)),\Phi((N_2,0))] &=& [(\sqrt{|\beta|}N_1,0),(\sqrt{|\beta|}N_2,0)]=
(0,|\beta|M_{12}^a)\nonumber\\
&=& \Phi((0,|\beta|M_{12}^a))= \Phi([(N_1,0),(N_2,0)]_{\beta})
\end{eqnarray}
where $M_{12}^a=q^{ab}(N_1\partial_bN_2-N_2\partial_bN_1)$ and we have more
specifically denoted the modified bracket by $[\cdot,\cdot]_{\beta}$ while
$[\cdot,\cdot]$ is the classical bracket. The anchor is preserved because
$Nn^{\mu}= \sqrt{|\beta|}\tilde{n}^{\mu}$ with a non-standard normal
$\tilde{n}^{\mu}$ such that
$g_{\mu\nu}\tilde{n}^{\mu}\tilde{n}^{\nu}=1/|\beta|$.  If $\beta$ is spatially
constant, as in models of first-order cosmological perturbations, modified
brackets of sections in the Lie algebroid $A$ are mapped to the classical
brackets on $A'$, with the required anchor because $N n^{\mu}\mapsto
(N/\sqrt{|\beta|})n^{\mu}=N \tilde{n}^{\mu}$. With spatially dependent
$\beta$, the existence of a morphism is less clear because $\{H[N],H_a[M^a]\}$
is not modified in effective models of loop quantum gravity, while it would
change in (\ref{vv}). Fiber transformations are therefore less general than
base transformations in mapping modified brackets to the classical ones.

The fiber map just introduced is valid only if $\beta$ has constant sign. When
$\beta$ is of indefinite sign, no $\beta$-absorbing morphism can exist: For
opposite signs of $\beta$, the corresponding groupoids are inequivalent
because their compositions are concatenations of slices in Lorentzian
space-time and 4-dimensional space of Euclidean signature, respectively.

For spatially constant $\beta>0$, we have a Lie algebroid morphism between
modified and unmodified brackets irrespective of where the deformation
function $\beta$ originates. In the modified case, we then have the classical
space-time structure after applying the morphism that absorbs $\beta$ in the
normalization. But the classical structure is obtained after a field
redefinition: The space-time metric obtained from $q_{ab}$ is not of the
standard canonical form but reads
\begin{equation} \label{dsN}
  {\rm d}s^2 = \epsilon\beta N^2{\rm d}t^2+ q_{ab} ({\rm
    d}x^a+N^a{\rm d}t)({\rm     d}x^b+N^b{\rm d}t)
\end{equation}
depending, in general, on $q_{ab}$ and $K_{ab}$. This line element is
conformally related to (\ref{dsbeta}).

\subsection{Equations of motion}

When we interpret hypersurface deformations as actual moves in space-time, we
refer to time-evolution vector fields, and therefore to coordinate
structures. Space-time coordinates are not quantized in canonical quantum
gravity, and therefore the vector field should not receive quantum corrections
if there is a classical manifold picture for the effective theory. Deformed
brackets with $\beta>0$ can sometimes be mapped to the classical space-time
structure in terms of hypersurface deformations, but this does not necessarily
lead to the same physics in terms of time evolution.

For a classical deformation with standard normalization, we use
\begin{equation}
\tau^{\mu}=\delta X^{\mu} = \delta N \tilde{n}^{\mu} + \delta N^{a} X^{\mu}_a
\label{eq:eqtau}
\end{equation}
in order to identify time deformations, while in the classical case with
non-standard normalization, we have
\begin{equation}
\delta X^{\mu} =
\delta N_{\beta}n^{\mu} + \delta N^{a} X^{\mu}_a
\end{equation}
with $n^{\mu}=\sqrt{|\beta|}\tilde{n}^{\mu}$.  These vector fields must be the
same: Changing the normalization of the normal vector should not affect the
relative position of two hypersurfaces $X^{\mu}$ and $X^{\mu}+\delta X^{\mu}$
embedded in space-time. Thus, the two time-evolution vector fields have to be
the same, and it follows that the infinitesimal lapse function $\delta
N_{\beta}$ of the modified theory must be given by
\begin{equation}
\delta N_{\beta} = \frac{1}{\sqrt{|\beta|}} \delta N\,.
\end{equation}

\subsubsection{Classical theory with non-standard normalization}

Classically, we have standard hypersurface-deformation brackets with the
normalization condition $g_{\mu\nu}n^{\mu}n^{\nu}=\epsilon$ and we know, by
\cite{Regained}, that second-order equations of motion for the metric are the
classical field equations of general relativity. However, we may change the
normalization condition to $g_{\mu\nu}n^{\mu}n^{\nu}=\epsilon|\beta|$.  The
theory is still classical, but the generator of normal deformations is
rescaled. Accordingly, the hypersurface-deformation brackets are
modified. Since the physics is insensitive to our choice of normalization, we
should be able to recover Einstein's field equations from the new brackets.

In \cite{Regained,LagrangianRegained} the Lie derivative with respect to the
normal vector field plays an important role in the derivation of possible
Hamiltonian constraints consistent with the brackets and hence in the
derivation of the equations of motion. One obtains a partial differential
equation which the Hamiltonian constraint as the generator of normal
deformations must obey \cite{Regained}, and similarly there is a related
partial differential equation for the Lagrangian \cite{LagrangianRegained}. If
the brackets are modified, the differential equation is changed by a new
coefficient $\beta$. For instance, a metric-dependent Lagrangian
$L[q_{ab}(x),K_{ab}(x)]$ consistent with constraints satisfying (\ref{HHbeta})
must satisfy the functional equation \cite{Action}
\begin{equation}
\frac{\delta L(x)}{\delta q_{ab}(x')}K_{ab}(x') + 2
(\partial_b\beta)(x)\frac{\partial L(x)}{K_{ab}(x)}\partial_a\delta(x,x')
+ 2 \beta(x)\frac{\partial L(x)}{\partial
  K_{ab}(x)}\partial_a\partial_b\delta(x,x') - (x\leftrightarrow x') = 0
\end{equation}
where $K_{ab}=\frac{1}{2}{\cal L}_nq_{ab}$ is taken with a non-standard normal
$n^{\mu}$.  The normal derivative is subsequently written as a Lie derivative
along $\tau^{\mu}$ in order to arrive at equations of motion with respect to
the time-evolution vector field. For the classical equations to result in this
second case, in which the algebroid and the normalization are modified in such
a way that we are still dealing with the classical theory, the function
$\beta$ appearing in $n^{\mu}$ with non-standard normalization (and therefore
in the Lie derivative $\mathcal{L}_{n}$ as well) must cancel the function
$\beta$ appearing in the modified brackets. We will make use of the presence
of such cancellations in our discussion of the modified case.

\subsubsection{Modified theory}

In models of loop quantum gravity, the hypersurface-deformation brackets are
modified. However, since one sets up the models in the standard canonical
formulation, the normalization $g_{\mu\nu}n^{\mu}n^{\nu}=\epsilon$ is
preserved. Since the normal does not depend on phase-space variables and is
not quantized, the normalization convention does not change. And yet, the
brackets are modified. This case is therefore different from simply rescaling
the normal vector. Nevertheless, one can understand the resulting structures
by rescaling the normal after new brackets have been obtained from quantum
effects. For spatially constant $\beta$, a morphism to the classical brackets
is obtained.  By applying the preceding arguments, we nevertheless expect
non-classical equations of motion: There is a function $\beta$ from the
modified brackets appearing in the Hamiltonian constraint or Lagrangian
regained from the brackets, but now there is no compensating $\beta$ in the
normal Lie derivative in relation to the $\tau^{\mu}$-derivative because it is
defined with respect to the standard normal vector $n^{\mu}$.

The dynamics is therefore modified, which is consistent with the results of
several detailed investigations of cosmological
\cite{ScalarGaugeInv,LoopMuk,InflConsist,InflTest,ScalarHolMuk,ScalarTensorHol,Omega,HolState,ObsLQC}
and black-hole consequences \cite{LTB,LTBII,ModCollapse,ModifiedHorizon} in
terms of physical, coordinate-independent effects.  An open question has been
whether one can introduce a modified effective space-time metric which is
generally covariant in the standard sense, or whether the deformed algebroid
modifies this symmetry and leads to an entirely new space-time structure.

For spatially constant $\beta$, we know that deformed brackets can be mapped
to classical brackets by a Lie-algebroid morphism so long as $\beta$ does not
change sign. In terms of space-time geometries, rescaling the normal vector
$n^{\mu}$ to $\tilde{n}^{\mu}=|\beta|^{-1/2}\,n^{\mu}$ then leads us back to
the unmodified brackets. We already know that this algebroid implements
standard space-time covariance in the canonical formalism. We therefore see,
in qualitative agreement with \cite{Absorb}, that a field redefinition allows
us to restore the undeformed brackets, and consequently general covariance in
the classical form.  The equations of motion are nevertheless different from
the classical ones because we moved the $\beta$ appearing in the modified
brackets into the new normal vector, which is not cancelled out when we
finally switch to equations of motion with respect to $\tau^{\mu}$.

\section{Consequences}

Hypersurface-deformation brackets can be modified by replacing the usual
normalization of the normal vector by
$g_{\mu\nu}n^{\mu}n^{\nu}=\epsilon\beta$, while the time-evolution vector
field must be the same for the modified as well as the unmodified
theory. These two facts raise the question of whether it is possible to
distinguish between classical modifications from non-standard normalizations
and modifications induced by quantum gravity theories. We have answered this
question in the affirmative because equations of motion with respect to a
fixed time-evolution vector field do change.

\subsection{Field equations and matter couplings}

If $\beta$ has definite sign and is spatially constant, one can absorb the
bracket modifications in a non-standard normalization. Gauge transformations
generated by the algebroid then amount to the standard symmetries of
covariance. Accordingly, regained constraints or Lagrangians must belong to
the canonical theory of some higher-curvature action, assuming that a local
effective action exists.

We expect higher-curvature effective actions when a local derivative expansion
exists. In canonical terms, a non-local quantum effective action is obtained
by coupling expectation values to independent quantum moments
\cite{EffAc,HigherTime}, which formally play the role of auxiliary fields in a
non-local theory. Only when moments behave adiabatically can they be
eliminated from the equations of motion, and a local effective action
results. As shown in \cite{EffConsQBR}, moments do not appear in structure
functions such as $\beta$ here, but they lead to higher-order constraints
which restrict the moments as independent variables. For a local,
higher-curvature version of the effective theory one would have to solve for
almost all the higher-order constraints, which may not always be possible. A
canonical effective theory still exists.

However, even if we have a standard higher-curvature effective action after a
field redefinition, there are additional effects from modified brackets. The
Hamiltonian constraint in such a system generates deformations along a
non-standard normal vector. Therefore, when equations of motion are written
with respect to a time coordinate, they belong to an effective action in which
time derivatives are multiplied by a factor of $\beta$. The main consequence
of modified algebroids is therefore a non-classical propagation speed, which
is in agreement with the specific results obtained in
\cite{ScalarGaugeInv,LoopMuk,ScalarHol,ScalarHolMuk,ScalarTensorHol,ScalarHolInv}
for cosmological scalar and tensor modes. From (\ref{dsN}), we have the
kinetic term $\ddot{\phi}/\beta-\Delta\phi$ in an equation of motion for a
scalar field on the effective Riemannian space-time. This result is in
agreement with a related one derived in \cite{Action} for metric-dependent
$\beta$, following \cite{LagrangianRegained,Action}.  At the same time, we
have generalized the result of \cite{Action} by extending it to functions
$\beta$ that may depend on extrinsic curvature as in cases of interest for
signature change.

One can turn these arguments around and try to generate explicit consistent
models with modified brackets by introducing non-standard normalizations in
different classical actions or constraints. More generally, we could relax the
orthogonality condition between $n^{\mu}$ and $X_a^{\mu}$ in order to find
models with the new modified brackets (\ref{NM}) and (\ref{NN}) with
$\alpha^a\not=0$. The recent analysis of \cite{RigidityGR} suggests that such
modified versions of constraints will have to be of higher than second order
in extrinsic curvature.

\subsection{(Non-)existence of an effective Riemannian structure}

Sometimes, the classical space-time structure is {\em assumed} in toy models
of quantum gravity, without checking closure of modified constraints. In fact,
one should not consider such constructions as models of quantum gravity but
rather of quantum-field theory on (modified) curved space-times because
quantum gravity is usually understood as including a derivation of
non-classical space-time structures in addition to a modified dynamics. For
instance, some constructions \cite{QFTCosmo,QFTCosmoClass,AAN} use
perturbation equations on a modified background $\bar{q}_{ab}$ subject to
evolution equations with quantum corrections. Perturbations are gauge-fixed or
combined into gauge-invariant expressions before quantization, and therefore one
assumes the classical space-time structure. As confirmed here, an effective
formulation with the classical space-time structure does exist as long as
$\beta>0$, but only after a field redefinition using either base
transformations or, in the case of a spatially constant $\beta$ as it is
realized in first-order cosmological perturbation theory, fiber
transformations of the hypersurface-deformation algebroid.

There are therefore two important caveats regarding assumptions as in
\cite{QFTCosmo,QFTCosmoClass,AAN}: First, if the evolution of $\bar{q}_{ab}$
is modified, a consistent description of space-time transformations for
inhomogeneous modes requires a modified $N$ which can only be computed if one
knows a consistent set of $\beta$-modified brackets. (The lapse function of
the postulated space-time metrics in \cite{QFTCosmo,QFTCosmoClass,AAN} do have
quantum corrections, but in an incomplete way that ignores the field
redefinition required for a consistent space-time structure.) The modified
$N$, as opposed to the classical $N$, then implies further quantum corrections
not directly present in the evolving $\bar{q}_{ab}$. One can, of course,
partially absorb $\sqrt{|\beta|}$ in $N'=\sqrt{|\beta|}N$ by introducing a new
time coordinate $t'$ with ${\rm d}t'=\sqrt{|\beta|}{\rm d}t$. But the
dependence of $\bar{q}_{ab}$ on this new $t'$ is different from the original
dependence on $t$, so that additional quantum corrections are present.

\subsection{Signature change}

In particular, as the second caveat, the signature of the effective space-time
metric can be determined only if one knows the sign $\epsilon\epsilon_{\beta}$
by which $\beta N^2{\rm d}t^2$ enters the metric (\ref{dsN}) in the equivalent
Riemannian space-time structure, which can differ from the classical value if
$\beta$ does not have definite sign. The sign, in turn, affects the form of
well-posed partial differential equations on the background; see for instance
\cite{Loss,SigImpl}. In the presence of signature change, there is no
deterministic evolution through large curvature. And even if one tries to
ignore this conclusion for a formal analysis of the resulting phenomenology,
no viable results are obtained \cite{QCExclude}.

If $\beta$ is of indefinite sign, it can no longer be absorbed globally. The
classical space-time structure can be used only to model disjoint pieces of a
solution in which $\beta$ has definite sign, corresponding to Lorentzian
space-time patches when $\beta$ is positive and Euclidean spatial patches when
it is negative. We then have non-isomorphic Lie algebroids.  A non-constant
sign of $\beta$ therefore triggers signature change
\cite{Action,SigChange,SigImpl} with the effective signature locally given by
$\epsilon\epsilon_{\beta}$. Globally, such a solution of an effective
quantum-gravity model can be described consistently only with a modified
algebroid, in which all structure functions are continuous and well-defined
even when $\beta$ goes through zero. It is no longer possible to absorb
$\beta$ globally, and therefore a new version of quantum space-time is
obtained.

\section*{Acknowledgements}

This work was supported in part by NSF grants PHY-1307408 and PHY-1607414.

\begin{appendix}

\section{ADM and geometrodynamics derivation of non-standard classical
  constraints}
\label{a:ADM}

We derive the results of Sec.~\ref{s:brief} for $\alpha^a=0$ using more
familiar methods.

\subsection{ADM}

Given a space-time metric $g_{\mu\nu}$ and a time-evolution vector field of
the form (\ref{eq:taudecompmod}) with respect to a foliation, we obtain the
canonical form of the metric by expanding $g_{\mu\nu}{\rm d}X^{\mu}{\rm
  d}X^{\nu}$ using
\begin{equation}
\mathrm{d}X^{\mu} = \partial_tX^{\mu}\mathrm{d}t
+ \partial_aX^{\mu}\mathrm{d}x^{a} = \left(N_{\beta}n^{\mu} +
  N^{a}X_a^{\mu}\right)\mathrm{d}t + X_a^{\mu}\mathrm{d}x^{a}
\end{equation}
with $N_{\beta}=N/\sqrt{|\beta|}$.  If $n^{\mu}$ has non-standard
normalization $g_{\mu\nu}n^{\mu}n^{\nu}=\epsilon\beta$, the metric
components are
\begin{equation} \label{CanonicalMetric}
g_{tt} = N^{a}N_{a} +\epsilon\beta N_{\beta}^2=
N^aN_a+\epsilon\epsilon_{\beta}N^2 \quad,\quad
g_{at} = N_a\quad,\quad
g_{tb} = N_b \quad,\quad
g_{ab} =q_{ab}\,.
\end{equation}

With respect to a non-standard normal, we define the tensor
\begin{equation}
K_{\mu\nu} = \frac{1}{2}\mathcal{L}_{n}q_{\mu\nu}\,.
\label{eq:extrinsic}
\end{equation}
It differs from the extrinsic-curvature tensor bu a factor of
$\sqrt{|\beta|}$, as can be seen from the alternative version
\begin{equation}
K_{\mu\nu} = \frac{1}{2N_{\beta}}\mathcal{L}_{\tau-\vec{N}}q_{\mu\nu}
\end{equation}
derived from (\ref{eq:extrinsic}) using (\ref{eq:taudecompmod}).  The
relationship between $K_{ab}= K_{\mu\nu} X_a^{\mu}X_b^{\nu}$ and the
$\tau$-derivative $\dot{q}_{ab}={\cal L}_{\tau}q_{ab}$ is therefore
\begin{equation}
K_{ab}=\frac{1}{2N_{\beta}}\left(\dot{q}_{ab}-\mathcal{L}_{\vec{N}}q_{ab}\right)\,.
\label{eq:Kab}
\end{equation}

In order to relate $K_{ab}$ to the momentum of $q_{ab}$, we need the
gravitational action $S=\int{\rm d}y^4 \sqrt{|\det g|}R$ in new variables
defined with respect to a non-standard normal. (We set $16\pi G=1$.) The
standard derivation from Gauss--Codazzi equations gives us the space-time
Ricci scalar
\begin{equation}
R = \mathcal{R} -\frac{\epsilon}{\beta}\left(K_{ab}K^{ab} - K^2\right)
\label{eq:Rsigma}
\end{equation}
expressed as a combination of the spatial Ricci scalar ${\cal R}$ and
$K_{ab}$.  (See also \cite{Embacher}, where a time-dependent $\beta$ has been
assumed to study classical signature change.)
Together with
\begin{equation}
\sqrt{|\det\left(X^{*}g\right)|} = N\sqrt{\det (g_{ab})}=
N_{\beta}\sqrt{|\beta|}\sqrt{\det\left(q_{ab}\right)}\,,
\label{eq:pulldet}
\end{equation}
all contributions to the Einstein--Hilbert action appear are written in terms
of new variables. The momentum of $q_{ab}$ is
\begin{equation}
P^{ab}(t,x) = \frac{\delta \mathcal{S}}{\delta\dot{q}_{ab}} =
-\frac{\epsilon\epsilon_{\beta}}{\sqrt{|\beta|}}
\sqrt{\det(q_{ab})}\left(K^{ab}-q^{ab}K_{c}^{c}\right)\,, \label{eq:PAB}
\end{equation}
while the momenta $P$ of $N$ and $P_a$ of $N^a$ vanish as usual. (The factor
of $\epsilon\epsilon_{\beta}/\sqrt{|\beta|}=(\epsilon/\beta)(N/N_{\beta})$ in
(\ref{eq:PAB}) is a result of combining $\epsilon/\beta$ in (\ref{eq:Rsigma})
with $N$ in (\ref{eq:pulldet}) and one of the $N_{\beta}$ obtained after
converting $K_{ab}$ to $\dot{q}_{ab}$ using (\ref{eq:Kab}).)  For the primary
constraints $P=0$ and $P_a=0$ to be preserved in time, we obtain as secondary
constraints the diffeomorphism and Hamiltonian constraints
\begin{eqnarray}
\mathcal{H}_a &:=& -2q_{ab} \nabla_bP^{bc} \label{eq:Ha}\\
\mathcal{H} &:=&-
\frac{\epsilon\epsilon_{\beta}\sqrt{|\beta|}}{\sqrt{\det(q_{ab})}}
\left(q_{ac}q_{bd}-\frac{1}{2}q_{ab}q_{cd}\right)P^{ab}P^{cd}
- \sqrt{|\beta|}\sqrt{\det(q_{ab})}\mathcal{R}\,. \label{eq:Hamconstr.}
\end{eqnarray}
These constraints have closed Poisson brackets corresponding to (\ref{vv}).
In terms of extrinsic curvature instead of the momentum, the first term of
(\ref{eq:Hamconstr.}) has a factor of $\epsilon_{\beta}/\sqrt{|\beta|}$, in
agreement with expressions regained from modified brackets \cite{Action}
following the methods of \cite{Regained,LagrangianRegained}.

\subsection{Geometrodynamics}

Using the formalism of hyperspace \cite{KucharHypI,KucharHypII,KucharHypIII},
the hypersurface-deformation brackets can be derived from infinitesimal
deformations, irrespective of the dynamics.  An infinitesimal deformation
$\delta X^{\mu}$ may be decomposed as
\begin{equation}
\delta X^{\mu} = \delta N_{\beta} n^{\mu} + \delta N^{a} X_{a}^{\mu}\,.
\label{eq:deltaX}
\end{equation}
The (non-standard) normalization and orthogonality relations
$g_{\mu\nu}n^{\mu}n^{\nu}=\epsilon\beta$ and $g_{\mu\nu}n^{\mu}X_{a}^{\nu}=0$
allow us to compute $\delta N_{\beta}$ and $\delta N^{a}$ from $\delta
X^{\mu}$:
\begin{equation}
\delta N_{\beta} = \frac{\epsilon}{\beta} n_{\mu}\,\delta X^{\mu}\quad , \quad
\delta N^{a} =
X^{a}_{\mu}\,\delta X^{\mu}
\label{eq:expressions}
\end{equation}
Here we do not refer to $\tau^{\mu}$ or $\delta N$ because the present
geometrical considerations refer to what is considered as the normal vector
with a non-standard normalization.

An arbitrary functional $F=F[X^{\mu}(x^{a})]$ on hyperspace changes if we
deform the hypersurface by $\delta N_{\beta}(x)$ along a normal geodesic and
stretch it by $\delta N^{a}(x)$. Using (\ref{eq:deltaX}), we write the
infinitesimal change of $F$ as
\begin{equation}
\delta F = \int_{\sigma}{\rm d}^3x \delta X^{\mu}(x)\frac{\delta}{\delta
    X^{\mu}(x)}F = \int_{\sigma}{\rm d}^3x
\left(\delta N_{\beta}(x)\rho_{0}(x) + \delta N^{a}(x)\rho_{a}(x)\right)F
\label{eq:deltaF}
\end{equation}
with the generators of pure deformations and pure stretchings
given by
\begin{equation} \label{rho0a}
\rho_0(x) := n^{\mu}(X(x))\frac{\delta}{\delta X^{\mu}(x)}\quad,\quad
\rho_a(x) := X^{\mu}_{a}(x)\frac{\delta}{\delta X^{\mu}(x)}\,.
\end{equation}
These generators can be interpreted as the Lie-algebroid anchor $\rho\colon
\Gamma(A)\to\Gamma(TB)$, with base manifold $B$ the space of embeddings
$X\colon \sigma\to {\cal M}$, expressed in a local basis: In a neighborhood
$U\subset B$, we introduce a smooth chart $(U,\{x^{a}\})$ of the manifold $B$
and a local frame $\{e_i\}$ for sections of the Lie algebroid
$\pi^{-1}(U)\subset A$. Then there exist smooth functions $c_{ij}^{k},
\rho_{i}^{a}\colon B\rightarrow \mathbb{R}$, such that
\begin{equation}
[e_i,e_j]_{A} = c_{ij}^{k}e_k \quad,\quad
\rho(e_i) = \rho_{i}^{a}\frac{\partial}{\partial x^{a}}\,.\label{eq:rhoia}
\end{equation}
These functions are called the structure functions of the Lie algebroid with
respect to the local frame $\{e_i\}$ and local coordinates $\{x^{a}\}$. For
the hypersurface-deformation algebroid, $\rho_0=\rho(e_0)$ and
$\rho_a=\rho(e_a)$.

There are infinitely many generators $\rho_0(x)$ and $\rho_a(x)$ which span
the tangent space to hyperspace at each hypersurface. Compared with the
coordinate basis $\delta/\delta X^{\mu}$, an important advantage of this basis
is its independence of the choice of space-time coordinates $X^{\mu}$. We can
therefore describe the kinematics in terms intrinsic to the
hypersurfaces. However, the basis is non-holonomic: commutators of the
generators $\rho_0(x)$ and $\rho_a(x)$ do not vanish in general.

In order to establish the commutators of deformation generators (\ref{rho0a})
we have to know how the normal vector changes under an infinitesimal
deformation.  To this end, the formula
\begin{equation}
\delta n^{\mu} = -\epsilon X^{\mu a}\delta N_{,a} + K_{ab} X^{\mu a}\delta
N^{b} - \Gamma^{\mu}_{\rho\sigma}X^{\rho}_{c}n^{\sigma}\delta N^{c} -
\Gamma^{\mu}_{\rho\sigma}n^{\rho}n^{\sigma}\delta N
\label{eq:deltan}
\end{equation}
has been used in \cite{Regained,KucharHypI} in order to compute the commutator
of normal deformations $\rho_0(x)$ in which $\delta n^{\mu}(x)/\delta
X^{\nu}(x')$ appear. Only the first term in (\ref{eq:deltan}) contributes to
this commutator, while all other terms are irrelevant for this purpose because
they present variations proportional to delta functions. Since delta functions
are symmetric in their arguments they will cancel out thanks to the
anti-symmetry of a commutator. The variation given by the first term in
(\ref{eq:deltan}), on the other hand, is proportional to
$\delta_{,a}(x,x')=-\delta_{,a'}(x',x)$, which is anti-symmetric and does
contribute.

The first term in (\ref{eq:deltan}) follows from a simple consideration that
can easily be extended to non-standard normalizations of $n^{\mu}$. One can
compute the full (\ref{eq:deltan}) in terms of its normal and tangential
components by varying $g_{\mu\nu}n^{\mu}n^{\nu}=\epsilon$ and
$g_{\mu\nu}X^{\mu}_an^{\nu}=0$. Since the first term in (\ref{eq:deltan}) does
not contribute to the normal component $n_{\mu}\delta n^{\mu}$, it must result
from $\delta(g_{\mu\nu}X^{\mu}_an^{\nu})=0$. This variation has three terms,
so that the equation can be solved for
\[
 X_{a\mu}\delta n^{\mu}= -n^{\mu}\delta X_{\mu,a}- X_a^{\mu}n^{\nu}\delta
 g_{\mu\nu}= -(n^{\mu}\delta X_{\mu})_{,a} -n^{\mu}_{,a}\delta X_{\mu}-
 X_a^{\mu}n^{\nu}\delta g_{\mu\nu}\,.
\]
The metric variations in the last term as well as $n^{\mu}_{,a}$ in the second
term can be written in terms of extrinsic curvature and the Christoffel
symbol, while the first term provides the first part of (\ref{eq:deltan}) upon
using (\ref{eq:expressions}) with $\beta=1$. For $\beta\not=1$, the first term
in (\ref{eq:deltan}) is replaced by $-\epsilon(\beta\delta N_{\beta})_{,a}$,
or $-\epsilon\beta(\delta N_{\beta})_{,a}$ if the derivative of $\beta$ is
combined with the last term in (\ref{eq:deltan}) which drops out of
commutators.  As a result, there is a factor of $\beta$ in the commutator
\begin{equation}
\left[\rho_0(x),\rho_0(x')\right] =
\epsilon\beta\left(q^{ab}(x)\,\delta_{,a}(x,x')\,\rho_{b}(x) -
  q^{ab}(x')\,\delta_{,a}(x',x)\,\rho_{b}(x')\right) \,.
\end{equation}
This result agrees with (\ref{vv}).

\end{appendix}


\end{document}